\documentclass[12pt]{article}
\usepackage{epsfig}

\newcommand{\be}{\begin{equation}}
\newcommand{\ee}{\end{equation}}
\newcommand{\ba}{\begin{eqnarray}}
\newcommand{\ea}{\end{eqnarray}}
\newcommand{\bi}[1]{\bibitem{#1}}
\def\aprle{\buildrel < \over {_{\sim}}}
\def\aprge{\buildrel > \over {_{\sim}}}

\begin{document}

\begin{titlepage}

\begin{center}
  
{\Large Tracing very high energy neutrinos from cosmological distances in ice}

\bigskip
\bigskip
 
\end{center}
\bigskip

\begin{center}
{J. Jones$^1$, I. Mocioiu${}^{1\mbox{,}3}$, M. H. Reno$^2$ and 
I. Sarcevic${}^{1\mbox{,}3}$}

\bigskip
\bigskip

{
$^1$ Department of Physics, University of Arizona, Tucson, Arizona 85721, 
USA \\
$^2$ Department of Physics and Astronomy, University of Iowa,
Iowa City, Iowa 52242 USA\\
$^3$ KITP, UCSB, Santa Barbara, CA 93106
}

\end{center}
\bigskip
\bigskip

\begin{abstract}
Astrophysical sources of ultrahigh energy neutrinos yield tau neutrino 
fluxes due to neutrino oscillations. We study in detail the contribution of 
tau neutrinos with energies above $10^6$ GeV relative to the contribution 
of the other flavors. We consider several different initial neutrino fluxes 
and include tau neutrino regeneration in transit through the Earth and 
energy loss of charged leptons. We discuss signals of tau neutrinos in 
detectors such as IceCube, RICE and ANITA. 

\end{abstract}

\end{titlepage}

\section{Introduction}

Very high energy neutrinos can be unique probes of both astrophysics and 
particle physics. They point back to their sources, escape from the 
most extreme environments and have energies much higher than those 
available in colliders. 

The SuperKamiokande experimental data on atmospheric neutrinos \cite{sk} shows
evidence of nearly maximal $\nu_\mu\leftrightarrow\nu_\tau$ mixing.
As a result, astrophysical sources of $\nu_\mu$
become sources of $\nu_\mu$ and $\nu_\tau$ in equal proportions after
oscillations over astronomical distances \cite{osc}. We do not differentiate 
between neutrinos and antineutrinos, since they can not be distinguished 
experimentally. 
The neutral current cross sections for  $\nu_\mu$ and 
$\nu_\tau$ are identical, and above $\sim 1 $ TeV the charged current 
cross sections are identical as well. 
Nevertheless, signals from $\nu_\tau$ have the potential
to contribute differently from $\nu_\mu$'s  because their propagation
through matter is different. Tau neutrinos produce
tau leptons via charged current interactions in the Earth.
Having a short lifetime, the tau leptons decay producing
$\nu_\tau$ which then interact and produce $\tau$, resulting in a
cascade that produces $\nu_\tau$ and
$\tau$ with energies lower than the original flux \cite{reg,reg1}. 
The leptonic tau decays also produce secondary $\nu_\mu$ and $\nu_e$ 
neutrinos \cite{secondarybeacom}.
All neutrinos from this cascade can then interact in the detector. 
The decays of 
taus in the detector also contribute to the signal \cite{dec,athar}. 
For muons, the electromagnetic energy loss coupled with the long
muon lifetime make regeneration of
 $\nu_\mu$ from muon decay negligible and high energy
$\nu_\mu$'s get large attenuation as they propagate through the Earth.

Signals of neutrino interactions in the rock below the ice or in ice
depend on the energy and flavor of the neutrino.
 Muon neutrino charged current (CC) conversions
to muons are noted by the Cherenkov signal of upward going muons in a
detector such as IceCube \cite{icecube}. High energy electromagnetic showers
from $\nu_e\to e$ CC interactions produce Cherenkov radiation
which is coherent for radio wavelengths \cite{askarian}.
The Radio Ice Cherenkov Experiment (RICE) has put limits on 
incident isotropic electron neutrino fluxes which produce downward-going 
electromagnetic showers \cite{rice}. The Antarctic Impulsive 
Transient Antenna (ANITA) also uses the ice as a neutrino converter 
\cite{anita}. These balloon missions will monitor the ice sheet
for refracted radio frequency signals with an effective telescope area of 
1M km$^2$. All flavors of neutrinos produce hadronic showers. In addition, 
tau decays contribute to both electromagnetic and hadronic showers that 
could be detected by IceCube, RICE or ANITA.

In this paper, we investigate the effect of $\nu_\tau$ regeneration from
tau decays and tau energy loss for neutrinos with energies above $10^6$ GeV,
with a particular interest in the higher energy range relevant for
RICE and ANITA. 
Attenuation shadows most of the upward-going solid angle for 
neutrinos, so we concentrate on incident neutrinos which are nearly horizontal
 or slightly upward-going. Considering the correct propagation of tau neutrinos
is important for a number of reasons. The effective volume for the detection 
of $\nu_\tau$ is larger than for the other flavors due to the regeneration 
effects. The initial interaction happens far outside the detector, but the
neutrinos (taus) produced in the neutrino interaction-tau decay cascade 
interact (decay) inside the detector. This can lead to enhanced event rates 
for particular energies and trajectories. The regenerated neutrinos and taus 
contribute, however, at lower energies than the initial ones, so a 
detailed quantitative analysis is necessary to understand where the 
regeneration effects are significant. Besides the possible enhancement in 
the rates, a good discrimination between all the different neutrino flavors is
important. The flavor composition of the detected neutrino fluxes could 
provide a better understanding of the sources of such neutrinos and
neutrino properties. The 
separate identification of each flavor could be important for this purpose
\cite{bq}.
 
We illustrate our results with a variety of fluxes. We concentrate
our analysis on the flux of GZK 
neutrinos \cite{gzk}. These are produced in the decay of pions from the 
interaction of cosmic ray protons with the background microwave photons, 
so they are a ``guaranteed'' source. In addition, they are expected to dominate
the neutrino flux at energies for which the radio Cherenkov detection methods
are most sensitive if the $Z$ burst model discussed below is not an
explanation of the highest energy cosmic rays.
Detection of GZK neutrinos 
could be essential for the understanding of very high energy cosmic ray
protons. 
The normalization of the GZK flux is somewhat uncertain. 
Our results are based on a conservative value, that considers 
``standard'' cosmological evolution \cite{gzk}. 
For ``strong'' source evolution the same 
general behavior of the flux is valid, but the flux is a factor of 4 higher, 
leading to higher rates. Other studies \cite{gzk2} obtain fluxes different by 
up to one order of magnitude. Since the normalization we use is the smallest, 
our results are conservative and rates could actually be higher. 

\begin{figure}
\epsfig{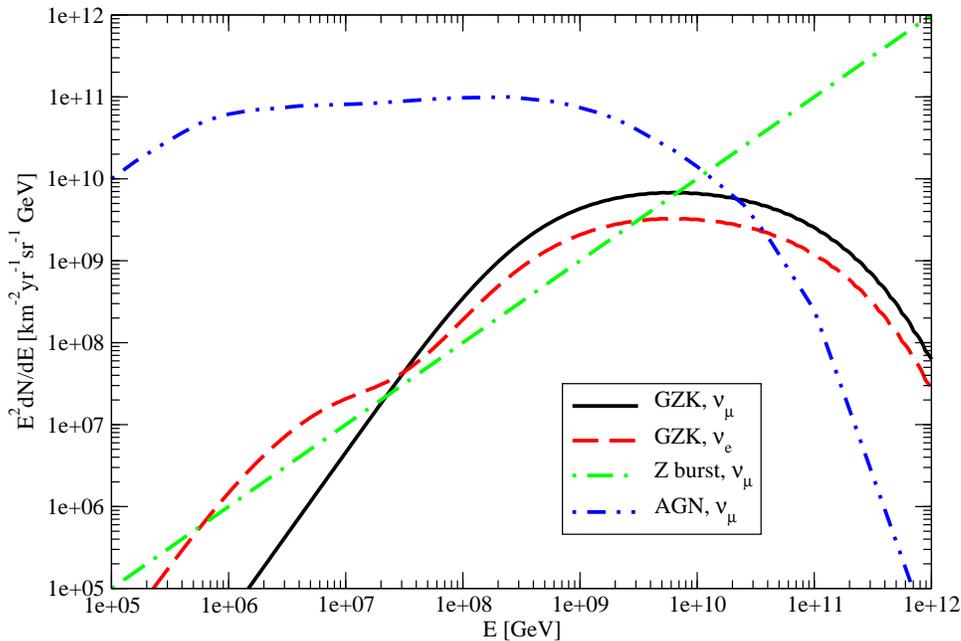}
\caption{Initial Neutrino Fluxes.}
\label{fig:fl}
\end{figure}

In order to discuss how results depend on the shape of the initial flux, 
we also show two generic distributions, $1/E$ and $1/E^2$. The $1/E$ flux 
describes well the neutrinos from $Z$ burst models \cite{sigl,zburst}.
$Z$ burst models could explain the highest energy cosmic rays by
extremely high energy neutrinos scattering on nearby massive
relic neutrinos. 
We also show results for the neutrino fluxes from AGN's for the model in 
Ref. \cite{mpr}. This flux has an
approximate $1/E^2$ behavior for $E=10^6-10^9$ GeV, so towards the lower end of
that energy range, our $1/E^2$ results apply.
All of the fluxes used in our analysis are shown in Fig.~\ref{fig:fl}, except for 
the $1/E^2$ distribution, which would correspond to just a horizontal line.

In the following section we describe our methods of computing the fluxes 
of neutrinos and charged leptons after propagation through the Earth. 
In section \ref{fl} we discuss the effect of regeneration and lepton energy 
loss on the propagation of the initial neutrino fluxes. In section \ref{sh}
we present our results for the distributions of electromagnetic and hadronic
showers from tau decays and neutrino interactions relevant to the different
types of experiments. Other recent papers \cite{dec,iceairsh} have 
focused on IceCube detection or on air showers. Related work has been 
recently presented at ICRC 2003 \cite{ICRC}. Here, we 
consider IceCube, as well as  
RICE and ANITA, which also use the ice as detector and have 
higher energy sensitivities. 
In the case of ANITA, one has dramatically longer trajectories through the 
ice to consider.

\section{Neutrino and Charged Lepton Propagation}

\begin{figure}[t]
\epsfig{file=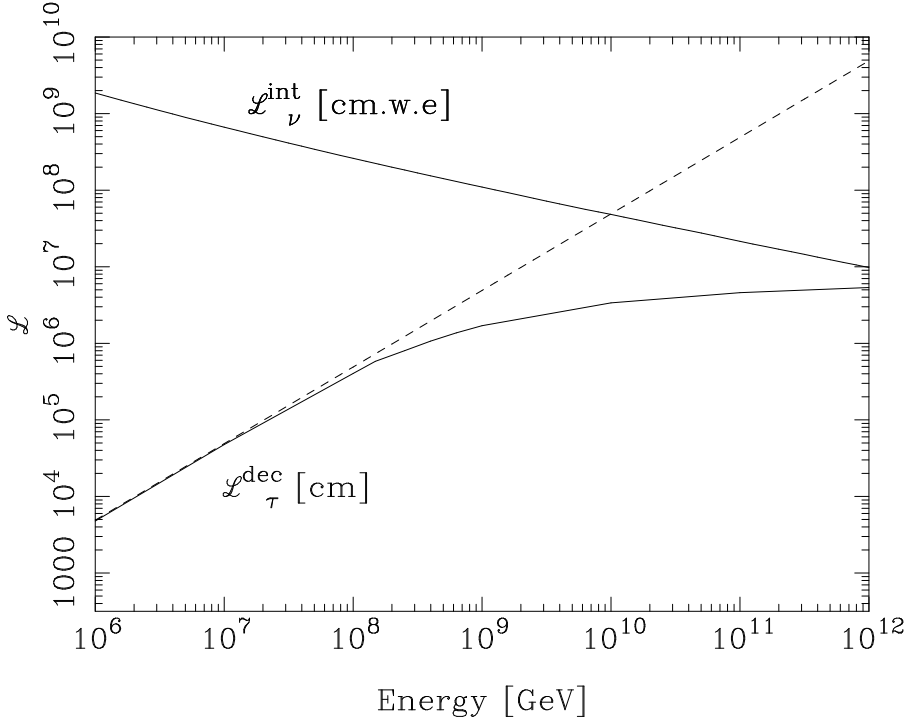,width=3.75in,angle=270}
\caption{Neutrino interaction length (solid line) and tau effective
decay length neglecting energy loss (dashed line) and including 
electromagnetic energy loss in water (solid line).}
\label{fig:intdec}
\end{figure}

Neutrino attenuation and regeneration are governed by interaction
lengths and decay lengths. In Fig. \ref{fig:intdec}, 
we show the neutrino interaction
length (in water equivalent distance), as well as the tau decay length
 and the effective decay length when one includes
tau energy loss in water \cite{gqrs,taudec}. 

The upper curve for the neutrino interaction length is equally
applicable to antineutrinos, since at high energies the
neutrino
and antineutrino cross sections with nucleons
are essentially equal because the cross sections are 
sea quark dominated. The neutral current contribution to the total 
cross section is about 1/2 of the charged current contribution.

To compare the interaction lengths with physical distances, we note
that $D=2 R_\oplus\cos\theta \rho\simeq 6 \times 10^8\
{\rm cm.w.e.}$ for $\theta=80^\circ $ and $D\simeq 2 \times 10^7\ 
{\rm cm.w.e.}$ for $\theta=89^\circ$ where $R_\oplus=6.37\times 10^8$ cm is 
the radius of the Earth. Neutrino attenuation is clearly an 
important effect, even for nearly horizontal incident neutrinos. In the figures
below, we mainly show results for a nadir angle of $85^\circ$ where attenuation
and regeneration effects in the propagation of $\nu_\tau$'s is
in effect without dramatically reducing the flux. We also compare fluxes with
those from incident $80^\circ$ and $89^\circ$ nadir angles. 

The effective decay length of the tau shows that, for energies
above about $10^8$~GeV, the tau is more likely to interact electromagnetically
than to decay \cite{taudec}. 
For an initial tau energy of $10^{12}$ GeV, the average
energy just before it decays is a ${\rm few}\times 10^8$ GeV, depending
on the density of the material the tau is passing through. Its 
effective decay length is of order 50 km in water. We use a
density of $\rho=0.9$ g/cm$^3$ for ice.

For tau neutrinos, we take into account the attenuation by charged current 
interactions, the shift in energy due to neutral current interactions and the
regeneration from tau decay. For tau leptons we consider their production in 
charged
current $\nu_\tau$ interactions, their decay, as well as 
electromagnetic energy loss.
 The tau neutrino and tau fluxes satisfy 
the following transport equations:
\ba
\frac{\partial F_{\nu_{\tau}}(E,X)}{\partial X}\!\! 
&\!\!\!\!=\!\!\!\!&\!\!-
N_A\sigma^{tot}(E) {F_{\nu_{\tau}}(E,X)}
+ N_A\int_E^\infty dE_y F_{\nu_{\tau}}(E_y,X)\frac{d\sigma^{NC}}{\!\!\!\!\!dE}
 (E_y,E)
\nonumber \\
&& + \int_E^\infty dE_y \frac{F_{\tau}(E,X)}{\lambda_\tau^{dec}}
\frac{dn}{dE}(E_y,E)
\label{nuprop}
\ea
\be
 \frac{\partial F_\tau(E,X)}{\partial X}=  
        - \frac{F_\tau(E,X)}{\lambda_\tau^{dec}(E,X,\theta)}
+ N_A 
\int_E^\infty dE_y F_{\nu_{\tau}}(E_y,X)\frac{d\sigma^{CC}}{\!\!\!\!\!dE} 
(E_y,E)
\label{tauprop}
\ee
\be
-\frac{dE_\tau}{dX}=\alpha+\beta E_\tau
\label{eloss}
\ee
Here 
$F_{\nu_{\tau}}(E,X)=dN_{\nu_\tau}/dE$ and $
 F_\tau(E,X)=dN_\tau/dE$ are the differential energy
spectra of tau neutrinos and taus
respectively, for lepton energy $E$, 
at a column depth $X$ in the medium defined by
\be
X = \int_0^L\rho(L')dL'.
\ee

We use the average density $\rho(L)$ of the medium along the neutrino path, 
as given by the Preliminary Earth Model \cite{prem}. We note that 
Antarctica is covered by a sheet of ice with a few km thickness, so that 
some of the neutrino trajectories can go mostly or even entirely through ice 
rather than rock. For the neutrino interaction cross-sections
we use CTEQ6 parton distribution functions \cite{CTEQ6} 
and their extrapolation in the 
regions of interest to us \cite{gqrs}. The decay length of the tau is 
$\lambda_{\tau}^{dec}(E,X,\theta)=\gamma c\tau_\tau\rho$. 
We use the decay distributions $dn/dE $ written explicitly in Ref. 
\cite{reg1},
and we approximate the energy distribution of the neutrino
interaction process with $d\sigma/dy(E,y)\simeq \sigma(E)\delta(y-\langle y
\rangle )$, with $y =(E-E')/E$ for incident
neutrino energy $E$ and outgoing lepton energy $E'$ and $\langle y \rangle
=0.2$. 

Equation (\ref{eloss}) 
describes the approximate
energy loss of the tau. The parameter $\alpha$ is determined
by the ionization energy loss and is negligible at high energy. 
The parameter $\beta$ is 
due to radiative energy loss through bremsstrahlung, pair production and 
photonuclear scattering. The photonuclear scattering becomes dominant at very 
high energies. As a first approximation we use 
$\beta=0.8\times 10^{-6}$ cm$^2$/g 
for the $\tau$ energy loss. There is a negligible 
change ($\aprle$ a few $\%$) if one uses a more realistic 
$\beta(E)= \bigl( 0.16+0.069\log_{10}(E_\tau/{\rm GeV})\bigr)
\times 10^{-6}$ cm$^2$/g \cite{taudec} except at very high energies 
($E_\tau > 10^{10}$ GeV), where the flux is extremely low. We use an energy loss 
parameter $\beta=7 \times 10^{-6}$ cm$^2$/g for the muons.

It should be added here that for $\bar\nu_e$ the additional scattering 
on electrons should be considered because of the W boson resonance at 
$6.3\times 10^6$~GeV. However, this resonance is extremely narrow (narrower 
than the energy resolution of the experiments) and it contributes a 
negligible amount to total rates \cite{athar, gqrs}. 

In \cite{secondarybeacom} it has been pointed out that 
the secondary fluxes of $\bar\nu_e$ and $\bar\nu_\mu$ from the tau decays 
could also have a significant contribution for a flux of mono-energetic neutrinos. 
In \cite{secondary} it has been shown however that
secondary neutrinos have negligible contribution for large nadir angles 
(i.e. for $\theta>60^\circ$) for generic initial neutrino fluxes, $1/E$ and $1/E^2$. 
The fluxes of secondary neutrinos are described by a transport equation similar to Eq. (\ref{nuprop}), 
with the decay distribution $dn/dE$ characteristic to each of the 
secondary neutrinos and we include them here, even though they are very small.

\section{Neutrino and Tau Fluxes \label{fl}}
	
>From Eqs. (\ref{nuprop},\ref{tauprop},\ref{eloss}) for $\tau$'s, and from
suitably modified equations for muons, we evaluate the charged lepton fluxes 
at the end of the trajectory of a neutrino incident with nadir angle
$\theta$. 

For most astrophysical sources, the neutrinos are produced in pion decays, which 
determine the flavor ratio $\nu_e:\nu_\mu:\nu_\tau$ to be $1:2:0$. After propagation over 
very long distances, neutrino oscillations change this ratio to $1:1:1$ because of the 
maximal $\nu_\mu\leftrightarrow\nu_\tau$ mixing. 
For the GZK flux, $\nu_e$ and $\nu_\mu$ incident
fluxes are different because of the additional contributions from 
$\bar{\nu}_e$ from neutron decay and $\nu_e$ from $\mu^+$ decays \cite{gzk}. Because of 
this, the flavor ratio at Earth is affected by the full three flavor mixing and is 
different from $1:1:1$. Given fluxes at the source 
$F^0_{\nu_e}$, 
$F^0_{\nu_\mu}$ and $F^0_{\nu_\tau}$, the fluxes at Earth become:
\ba
F_{\nu_e}&=&F^0_{\nu_e}-\frac{1}{4}\sin^22\theta_{12} (2 F^0_{\nu_e}-F^0_{\nu_\mu}-
F^0_{\nu_\tau})
\label{fle}
\\
F_{\nu_\mu}&=&F_{\nu_\tau}=\frac{1}{2}(F^0_{\nu_\mu}+F^0_{\nu_\tau})+\frac{1}{8}\sin^22\theta_{12}
(2 F^0_{\nu_e}-F^0_{\nu_\mu}-F^0_{\nu_\tau})
\label{flmutau}
\ea
where $\theta_{12}$ is the mixing angle relevant for solar neutrino oscillations. We 
have assumed that $\theta_{23}$, the mixing angle relevant for atmospheric neutrino 
oscillations, is maximal and $\theta_{13}$ is very small, as shown by reactor experiments,
as well as atmospheric and solar data. 
We use the initial GZK flux evaluated by Engel, Seckel and 
Stanev \cite{gzk} for standard evolution and we get the fluxes at Earth from the equations
(\ref{fle}) and (\ref{flmutau}). The fluxes of $\nu_\mu$ and $\nu_\tau$ are still equal, 
due to the maximal $\nu_\mu\leftrightarrow\nu_\tau$ mixing. 
The main effect of the three flavor oscillations is to 
transform some of the low energy $\nu_e$'s into $\nu_\mu$ and $\nu_\tau$. 
This could be 
useful since IceCube has very good sensitivity for detecting tracks, which are 
enhanced in this case. 

\begin{figure}[t]
\epsfig{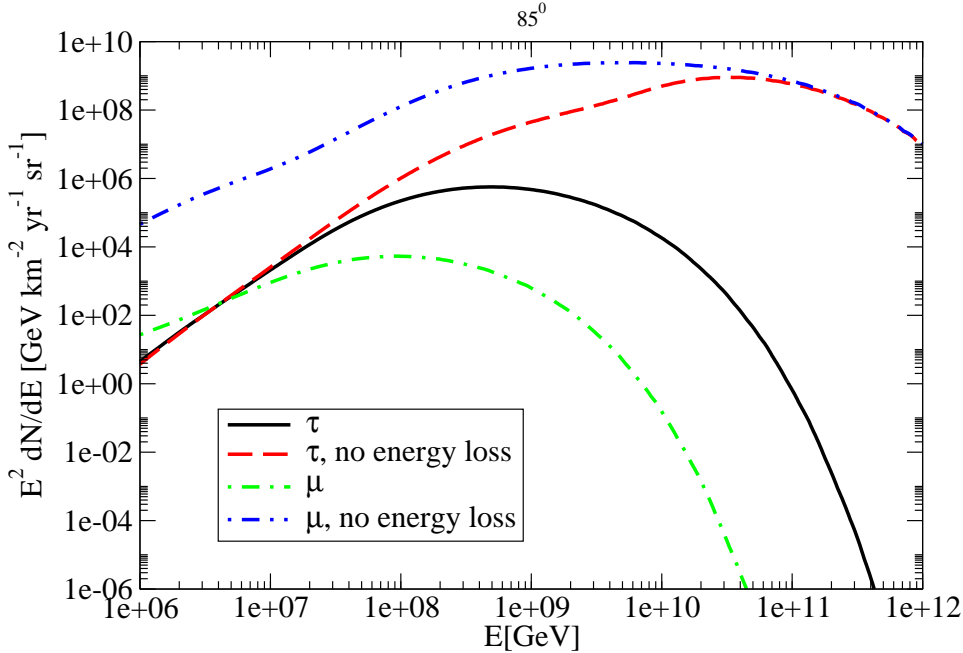}
\caption{Charged lepton distributions for GZK neutrinos, at a nadir angle
of $85^\circ$.}
\label{fig:tau}
\end{figure}

To illustrate the effect of the charged lepton lifetime and electromagnetic
energy loss in propagation, we begin with muon and tau distributions.
Fig. \ref{fig:tau} shows the tau and muon distributions for the GZK neutrinos
 at the end of the trajectory through the Earth with a nadir angle of 
$85^\circ$, with and without taking into account the
energy loss of the charged lepton. As noted above, the incident $\nu_\mu$ and
$\nu_\tau$ fluxes are equal.
It is clear from here that the energy loss is extremely important 
and strongly limits the fluxes at very high energy. 
Including energy loss, the difference between the $\tau$ and $\mu$
flux above $10^7$ GeV comes from the $\nu_\tau$ pileup from interaction and 
regeneration, as well as to the difference in decay length and energy loss 
for taus and muons. Most of the charged leptons are produced
in the last step of neutrino propagation. The muon flux below $10^7$ GeV
dominates the tau flux because of the much longer decay 
length of the muon. 

\begin{figure}[t]
\epsfig{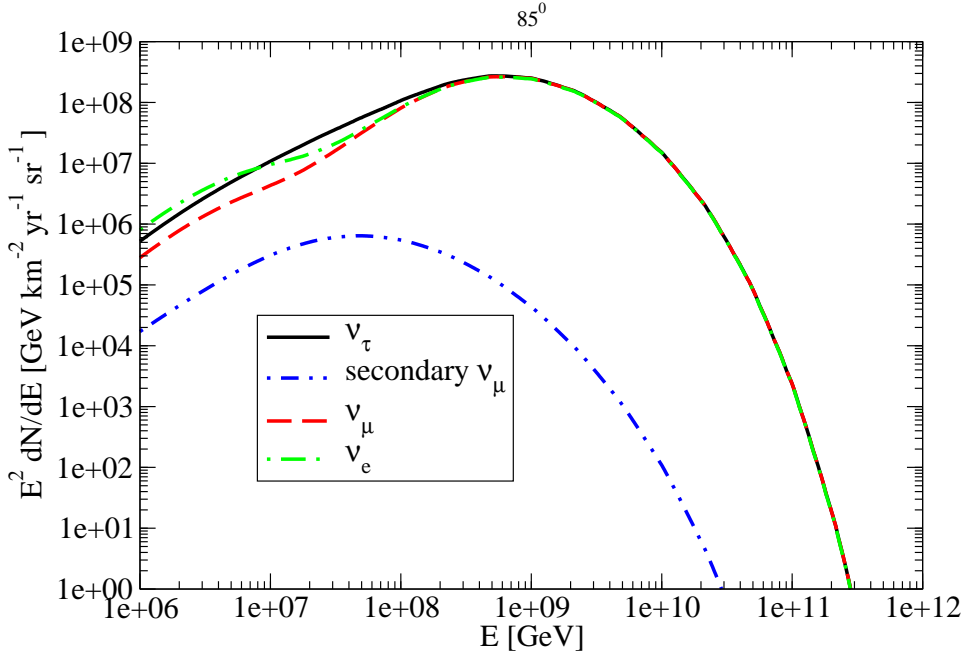}
\caption{Neutrino distributions for GZK neutrinos, at a nadir angle
of $85^\circ$.  }
\label{fig:nu}
\end{figure}

Fig. \ref{fig:nu} shows the distribution of $\nu_\tau$, $\nu_\mu$ 
and $\nu_e$ fluxes after propagation through the Earth 
for the same initial flux and angle. For $\nu_e$ and
$\nu_\mu$ the transport equations are effectively decoupled from those of 
the charged leptons and their propagation is given by Eq.~(\ref{nuprop}) in 
which the last term is not present. On the same figure we show the flux of 
secondary $\nu_\mu$ neutrinos, which is the same with that of the secondary 
$\nu_e$. It can be seen that their contribution is negligible, with maximum 
corrections of the order of a few percent around $10^7$ GeV and much smaller
for most energies. At larger nadir angles, the secondary neutrino contribution
will be an even smaller fraction of the $\nu_e$ and $\nu_\mu$ fluxes. At smaller
nadir angles, the secondary flux will be a larger percentage of the overall
neutrino flux, but the flux will become more strongly attenuated.

\begin{figure}[t]
\epsfig{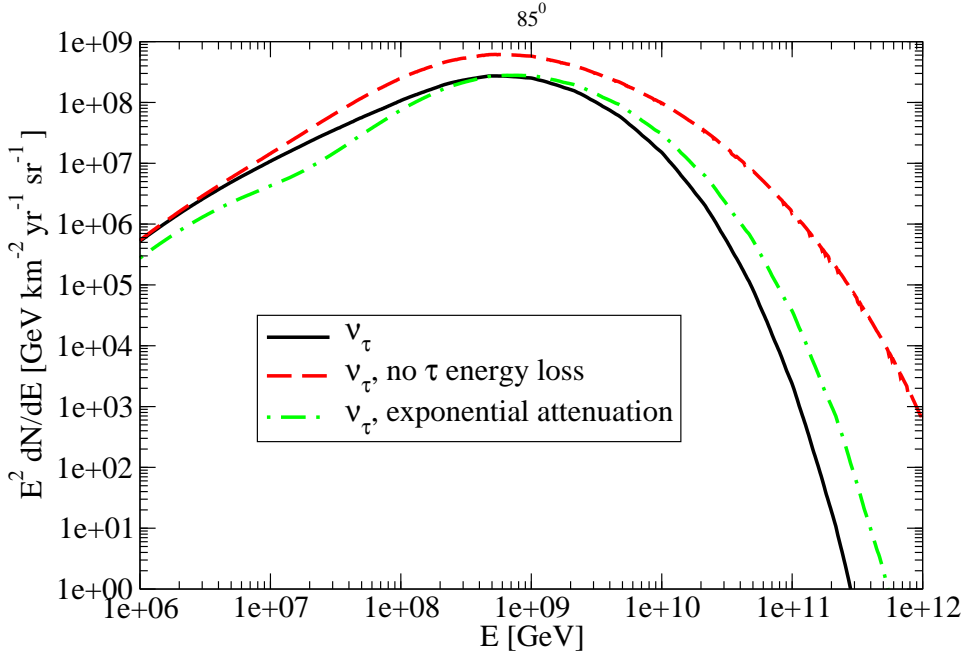}
\caption{Tau neutrino distributions for the GZK  flux, at a nadir angle
of $85^\circ$.  }
\label{fig:nuc}
\end{figure}

\begin{figure}[t]
\epsfig{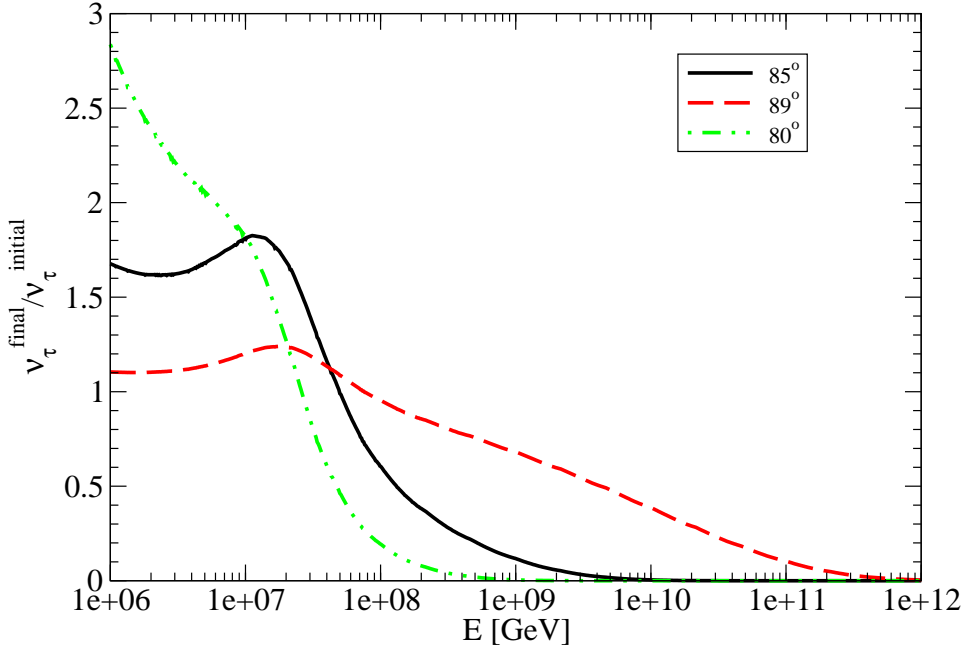}
\caption{Ratio $\nu_\tau^{\rm final}/\nu_\tau^{\rm initial}$ for GZK neutrinos, 
at nadir angles of $80^\circ$, $85^\circ$ and $89^\circ$.}
\label{fig:ratio}
\end{figure}

Fig. \ref{fig:nuc} shows the same $\nu_\tau$ distribution as in the previous
figure, together with the distributions obtained if energy loss of tau 
leptons or the shift in energy due to neutral current is not included in 
the propagation. Electromagnetic energy loss is extremely important
at energies above $10^8$ GeV and this effect causes strong
suppression of the neutrino fluxes at very high energy. If the tau would not 
lose energy in its propagation, the regeneration effect would be much bigger 
at high energy. 
The $\nu_\tau\to \tau\to \nu_\tau$ regeneration is a significant effect for
energies between $10^6$ and $10^8$ GeV, as can be seen more
clearly in Fig.~\ref{fig:ratio}. The neutral current interactions
of the neutrinos are also very important, as can be seen 
in Fig. \ref{fig:nuc} by comparing the flux of $\nu_\tau$ after correct 
propagation with a simple attenuation with 
$exp (-D/{\cal L}^{\nu}_{int}(E))$ for column depth $D$ and
interaction 
length ${\cal L}$ evaluated using the charged current cross section.
The difference can be as large as two orders of magnitude at very high energy.

\begin{figure}[t]
\epsfig{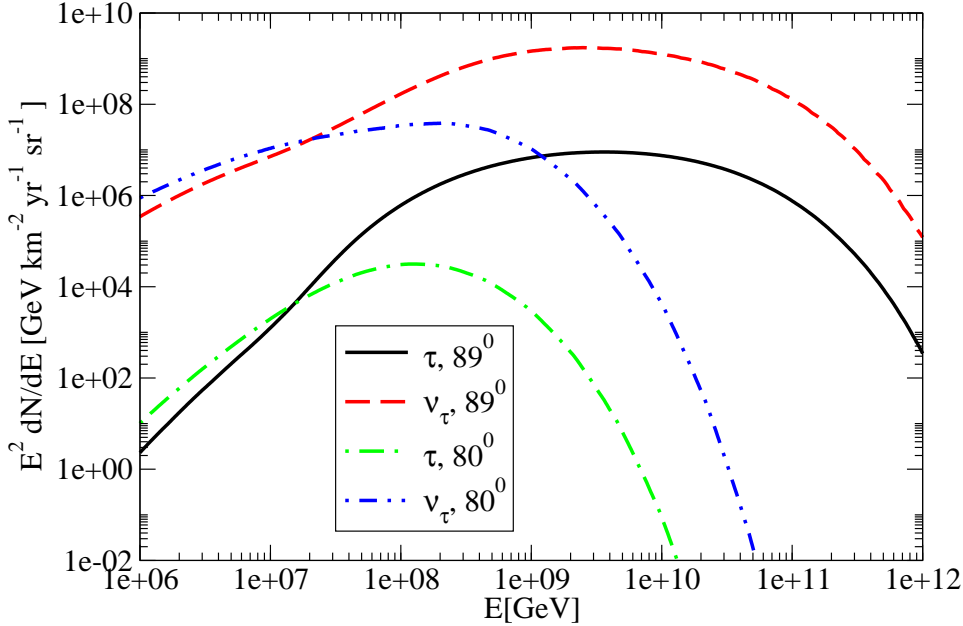}
\caption{The
$\tau$ and $\nu_\tau$ distributions for GZK neutrinos for different
nadir angles. }
\label{fig:angles}
\end{figure}

Fig. \ref{fig:angles} shows the $\tau$ and $\nu_\tau$ distributions for 
the same initial neutrino flux, but for different nadir angles.  
As the neutrinos pass through more matter, the regeneration effects become 
relatively more important, but the overall fluxes get significantly 
attenuated. 
At a nadir angle of $80^\circ$ the Earth is already opaque to neutrinos with 
energies above $10^{10}$ GeV. 
	
Fig. \ref{fig:ratio} shows the ratio of the tau neutrino flux after
propagation  to incident tau neutrino flux, for $89^\circ$, $85^\circ$and $80^\circ$.
This ratio illustrates a combination of the 
regeneration of $\nu_\tau$ due to tau decay and the attenuation of all 
neutrino fluxes. 
For $89^\circ$, where both the total distance and the density are smaller, 
the attenuation is less dramatic, and the flux can be significant
even at high energy. The regeneration in this case can add about $25\%$ 
corrections at energies between $10^7$ and $10^8$ GeV. 
For $85^\circ$ the relative enhancement is around $80\%$ and peaked at 
slightly lower 
energies, while at $80^\circ$ it is almost a factor of 3 at low energy. At 
$80^\circ$, however, the flux is very strongly attenuated for energies above 
a few $\times 10^7$ GeV. 

It is already clear from Fig. \ref{fig:ratio} 
that the total rates will be dominated by the 
nearly horizontal trajectories 
that go through a small amount of matter. The largest pileups occur when the
trajectory of the neutrino passes through a larger column depth. For the
higher energies relevant to RICE and ANITA, this doesn't translate to
higher fluxes of $\nu_\tau$'s. Attenuation is the main effect at those energies.
Rates can get significant 
enhancements at low energies where the regeneration from tau decays adds an 
important contribution even for longer trajectories. 

\section{Showers\label{sh}}

The above distributions illustrate the effects of the propagation 
through matter on the initial neutrino fluxes. We now turn to the 
experimentally relevant observables, which are the showers.

We consider hadronic showers produced in $\nu_\tau$ interactions, as well
as electromagnetic and hadronic showers from $\tau$ decays in ice. All other 
flavors of neutrinos will contribute to the hadronic showers, while $\nu_e$ 
will also contribute to electromagnetic showers. The shower distributions 
will be different for each type of experiment, depending on the detector 
setup. 

We start by considering contributions to electromagnetic and hadronic showers 
in a depth of $\sim 1$~km of ice, relevant for experiments like IceCube and 
RICE. 

We use Eqs. (\ref{nuprop}-\ref{eloss}) and the corresponding ones for $e$ and 
$\mu$ flavors to obtain the tau and neutrino distributions 
($F_\tau(E_\tau)$ and $F_\nu(E_\nu)$)
after propagation up to the last few km of ice. 
We then estimate the shower distributions that are produced over the distance 
$d\sim$ few km. The contributions of neutral currents and regeneration over
a few kilometers is small and here we neglect them for the last part of
the trajectory, which is where the showering takes place. This can no longer 
be done when showers are detected over long distances, as will be the 
case for ANITA.

There are two contributions to electromagnetic showers from tau decays.
One is from the decay of the taus produced outside the detector that 
decay electromagnetically inside the detector, giving ``lollipop'' events 
in IceCube and electromagnetic showers in RICE. The distribution of these 
showers is given as a function of the energy of the electron produced in the 
decays by:
\ba
F_{sh,\tau}^{em,1}(E_e)&=&
 \int_{E_e}^{E_{max}} dE_\tau \int_0^{d\rho}dx F_\tau(E_\tau) 
\nonumber \\
& &e^{-x/\lambda_\tau^{dec}(E_\tau e^{-\beta x})}\frac{1}
{\lambda_\tau^{dec}(E_\tau e^{-\beta x})}
\frac{dn_{\tau\to e}}{dE_e}(E_\tau e^{-\beta x},E_e)
\ea
 where $F(E)\equiv{dN}/{dE}$ is the flux of taus just before 
the final column depth $d$ in ice and 
$\rho=0.9$ g/cm$^3$ is the density of ice.
For negligible energy loss (which is a very good approximation below 
$10^9$~GeV in 1 km of ice) this becomes:
\be
F_{sh,\tau}^{em,1}(E_e)=\int_{E_e}^{E_{max}} dE_\tau 
F_\tau(E\tau)
\left( 1-e^{-d\rho/\lambda_\tau^{dec}(E_\tau)}\right )
\frac{dn_{\tau\to e}}{dE_e}(E_\tau,E_e)
\label{emsht1}
\ee
The other contribution comes from the taus that are produced and decay
electromagnetically inside the detector. In IceCube these correspond to 
``double bang'' events. The distribution is
obtained from:
\ba
F_{sh,\tau}^{em,2}(E_e)&=&\int_0^{d\rho}dx\int_x^{d\rho}dy
\int_{E_e}^{E_{max}}dE_{\nu_\tau} \int_{E_e}^{E_\nu} dE_\tau \,\, 
F_{\nu_\tau}(E_{\nu_\tau}) \nonumber \\
& & e^{-xN_A\sigma^{CC}(E_\nu)}
N_A\frac{d\sigma^{CC}}{dE_\tau}(E_{\nu_\tau},E_\tau)  \nonumber \\
& & e^{-y/\lambda_\tau^{dec}(E_\tau e^{-\beta y})}
\frac{1}{\lambda_\tau^{dec}(E_\tau e^{-\beta y})}
\frac{dn_{\tau\to e}}{dE_e}\left({E_e},{E_\tau e^{-\beta y}}\right)
\ea
When energy loss is negligible, this becomes:
\ba
F_{sh,\tau}^{em,2}(E_e)\!\!\!\!\!& &=
\int_{E_e}^{E_{max}}dE_{\nu_\tau} \int_{E_e}^{E_\nu} dE_\tau \,\, 
F_{\nu_\tau}(E_{\nu_\tau}) 
N_A\frac{d\sigma^{CC}}{dE_\tau}(E_{\nu_\tau},E_\tau) 
\frac{dn_{\tau\to e}}{dE_e}\left({E_e},{E_\tau}\right) \nonumber \\
& & \Bigg( 
\frac{1}{N_A\sigma^{CC}(E_{\nu_\tau})+1/\lambda_\tau^{dec}(E_\tau)}
 \left( 1-e^{-d\rho(N_A \sigma^{CC}(E_{\nu_\tau})+1/\lambda_\tau^{dec}
(E_\tau))}\right )-\nonumber\\
& & \frac{1}{N_A \sigma^{CC}(E_{\nu_\tau})}
e^{-d\rho/\lambda_\tau^{dec}(E_\tau)}
\left( 1-e^{-d\rho N_A \sigma^{CC}(E_{\nu_\tau})}\right ) \Bigg)
\label{emsht2}
\ea

The main source of electromagnetic showers comes from the $\nu_e$ 
charged current interactions given by: 
\be
F_{sh,\nu}^{em}=\int_{E_e}^{E_{max}}
 dE_\nu F_{\nu}(E_{\nu})(1-e^{-d\rho N_A \sigma^{CC}(E_\nu)})
\frac{1}{\sigma^{CC} (E_\nu)}\frac{d\sigma^{CC}}{dE_e}(E_\nu,E_e)
\label{emshnu}
\ee

Hadronic showers from tau decays are given by Eqs.(\ref{emsht1})
with the electron replaced by hadrons.

The hadronic showers from $\nu_\tau$ interactions are given by:
\be
F_{sh,\nu}^{h}=\int_{E_h}^{E_{max}}
 dE_\nu F_{\nu}(E_{\nu}) 
(1-e^{-d\rho N_A \sigma (E_\nu)})
\frac{1}{\sigma (E_\nu)}\frac{d\sigma}{dE_h}(E_\nu,E_h)
.
\label{nuhsh}
\ee
In Eq.~(\ref{nuhsh}), $\sigma$ is the total cross section and the 
hadronic energy comes from the energy transfer to the target nucleus.

Hadronic showers are also produced in $\nu_\mu$ and $\nu_e$ 
neutral current and charged current interactions.
These are also obtained from Eq.~(\ref{nuhsh}), using the corresponding fluxes.

We briefly describe below all types of interactions and their 
signatures in the various experiments. We then show shower distributions 
relevant in each case.

\subsection{Showers in kilometer size detectors: IceCube, RICE}

A ${\rm km}^3$ Cherenkov detector in ice, IceCube \cite{icecube} has 
sensitivity to high energy neutrinos up to $\sim 10^8-10^9$ GeV, the energy 
range where distinguishing 
the effects of tau neutrinos from muon and electron neutrinos could be 
possible.


Neutral current interactions of all types of neutrinos produce hadronic 
showers. 
Charged current interactions of $\nu_e$ contribute to both hadronic and 
electromagnetic showers. It is likely that IceCube would observe a big 
shower with the total energy of the incoming neutrino rather than be able
to separate the two components (hadronic and electromagnetic). 
Charged current interactions of $\nu_\mu$ give a hadronic shower and a muon. 
The muon track is the main signal in IceCube, so these events are easily 
identifiable, containing a shower and a long track emerging from it.
At high energies however, the same signal could be produced by a $\nu_\tau$
interaction, since the decay length of the tau becomes longer than the size 
of the detector for energies above  a few$\times 10^7$~GeV.

Charged current $\nu_\tau$ interactions can have very different signatures
depending on energy. At $10^6$ GeV, the tau decay length is $\sim$50 m, 
and the shower (hadronic or electromagnetic) from the tau decay cannot be 
separated from the hadronic shower from the initial $\nu_\tau$ interaction. 
At a few times $10^6$ GeV the range of the tau becomes a few hundred 
meters and can give the characteristic signal of the ``double bang'' events
\cite{db}. 
These are events where the shower from the neutrino interaction and the shower
from the tau decay can be separated and are both observed in the detector, 
together with the tau track. At a few$\times 10^7$ GeV the tau decay length 
is already longer than 1 km, and the taus look like muons. What is seen 
is the shower from the neutrino interaction and then a track. 

Taus produced in neutrino interactions outside the detector can generate 
``lollipop'' events. The initial shower from the neutrino interaction that 
produces the tau is missed and what is seen is the track of the tau and the 
shower from its decay that IceCube can identify as electromagnetic or 
hadronic from the existence of muon tracks in the shower.

RICE \cite{rice} uses dipole antennas in the Antarctic ice to measure radio 
frequency Cherenkov radiation from high energy showers. For energies around 
$10^9$~GeV the effective volume of the detector is 
$\sim 15\ {\rm km}^3 {\rm sr}$. At present, RICE limits the fluxes of 
downward $\nu_e$ with energies between few$\times 10^7$ and $10^{12}$ GeV 
\cite{rice}. 
Interactions modeled by a Monte Carlo at an energy of $10^9$~GeV are detected
from within a depth of 1 km from the surface and out to a radial distance of 
about 4 km. The effective volume of the detector has a strong dependence on 
energy below $10^9$ GeV. Hadronic and electromagnetic showers can be separately
identified, with somewhat different effective volumes, depending on energy.
The RICE experiment can therefore measure electromagnetic showers from 
$\nu_e$ interactions, hadronic showers from the charged and neutral
current interactions of all flavors of neutrinos and electromagnetic and 
hadronic showers from tau decays.

\begin{figure}[t]
\epsfig{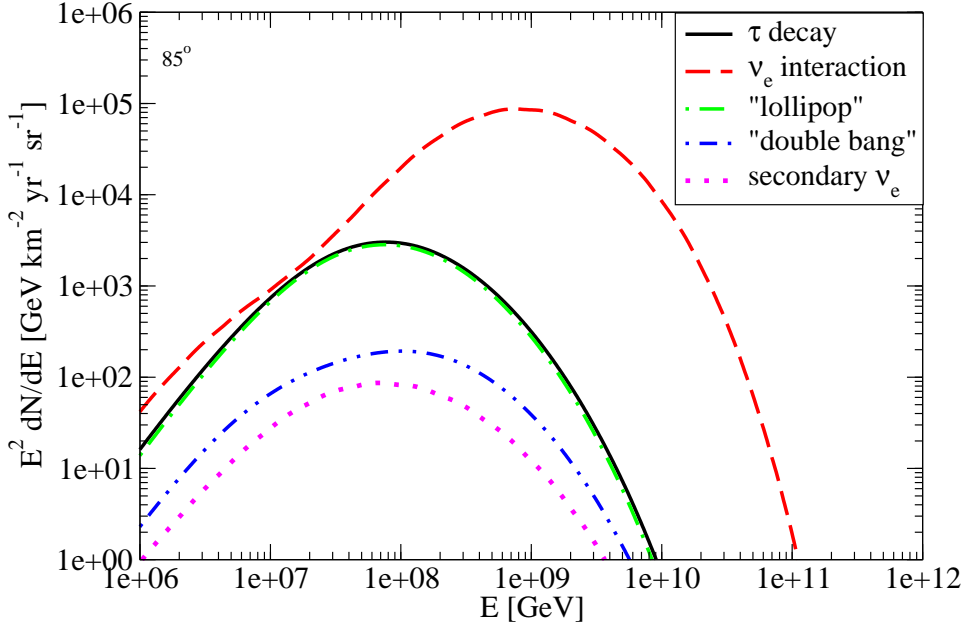}
\caption{Electromagnetic shower distributions for GZK neutrinos, 
at a nadir angle
of $85^\circ$ for a km size detector. The $\tau$ decay curve is the sum 
of the lollipop and double bang curves. }
\label{fig:emsh}
\end{figure}

Fig. \ref{fig:emsh} shows the electromagnetic shower distributions at a 
nadir angle of 85$^\circ$. In the 
absence of the tau neutrinos, only $\nu_e$ interactions lead to electromagnetic
showers. These $\nu_e\rightarrow e$ CC conversions 
still dominate at high energies, even when tau neutrinos 
produce tau leptons that decay electromagnetically. At lower energies however, 
the taus add a significant contribution to the electromagnetic shower
signal. Tau decays give their most important relative contribution to
electromagnetic showers at electron energies of a few$\times 10^7$GeV.
These decays are separated in ``double bang'' and ``lollipop'' events on the
same plot. It can be seen that the contribution of the ``double bang'' is 
relatively small. They do not contribute significantly to total
event rates, but even in very small numbers they are important as a 
characteristic signature for taus. Secondary $\nu_e$ interactions also 
produce electromagnetic showers, but it can be seen that their contribution 
is very small.

\begin{figure}[t]
\epsfig{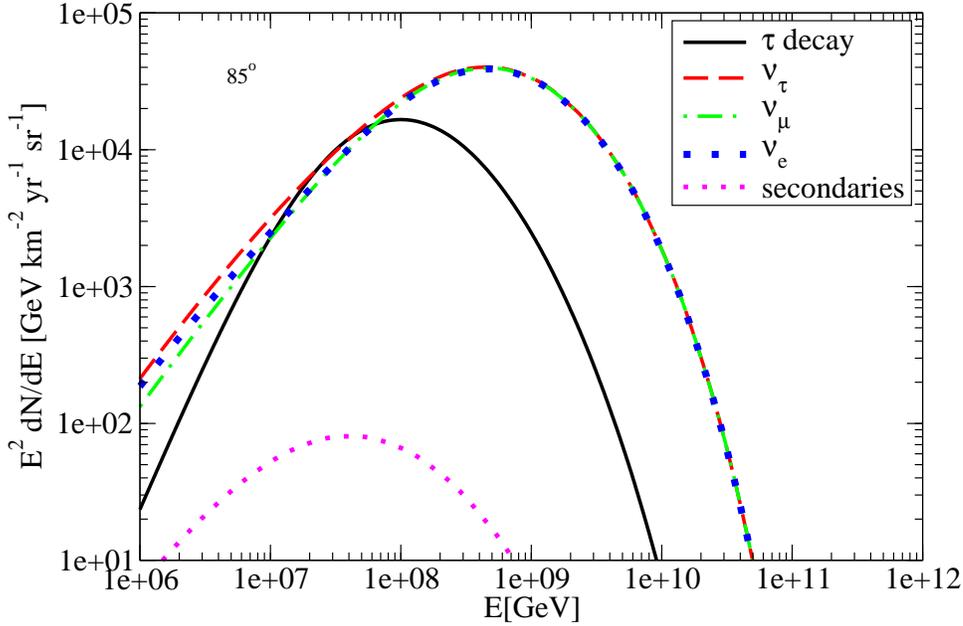}
\caption{Hadronic shower distributions for GZK neutrinos, 
at a nadir angle of $85^\circ$ for a km size detector.}
\label{fig:hsh}
\end{figure}

Fig. \ref{fig:hsh} shows the hadronic shower distributions. The features seen 
in the neutrino fluxes are recovered in the hadronic showers as well. The
shower rates from $\nu_\tau$, $\nu_\mu$  and $\nu_e$ are 
nearly identical at shower energies above $10^8$ GeV. The tau neutrino 
pileup is important below $10^8$~GeV. Between $10^7$ and $10^8$~GeV
tau decays give the main contribution to the hadronic shower rate.
Showers from $\nu_e$ and $\nu_\mu$ secondary neutrinos are also shown and 
their contribution is negligible.

\begin{figure}[t]
\epsfig{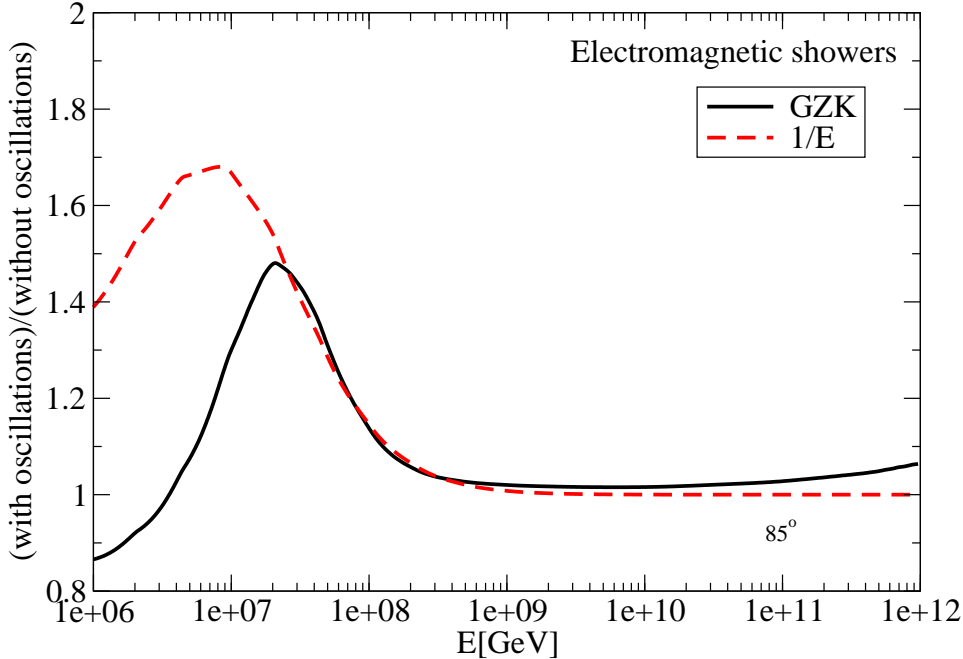}
\caption{Ratio of electromagnetic shower rates in the presence
and absence of 
$\nu_\mu\to\nu_\tau$ oscillations for GZK and $1/E$ neutrino spectra 
for nadir angle $85^\circ$ for a km size detector.}
\label{fig:shratioem}
\end{figure}

Fig. \ref{fig:shratioem} shows the ratio of the electromagnetic 
shower rates at nadir angle $85^\circ$ in the presence and absence of 
oscillations for the GZK and $Z$ burst 
neutrino fluxes. This ratio illustrates the effect of the oscillations 
on the signal and in particular the possible enhancements due to tau pileup. 
In absence of oscillations, the only contribution to 
electromagnetic showers comes from $\nu_e$ interactions. In the presence of 
$\nu_\mu\to\nu_\tau$ oscillations, electromagnetic decays of taus from tau 
neutrinos add significant contributions to these rates at energies below 
$10^8$ GeV. In the same time, for the GZK flux, $\nu_e\to\nu_{\mu,\tau}$ 
oscillations
reduce the number of $\nu_e$'s at low energy, such that below a few 
$\times 10^6$ GeV there are fewer electromagnetic showers than in the absence 
of oscillations.

\begin{figure}[t]
\epsfig{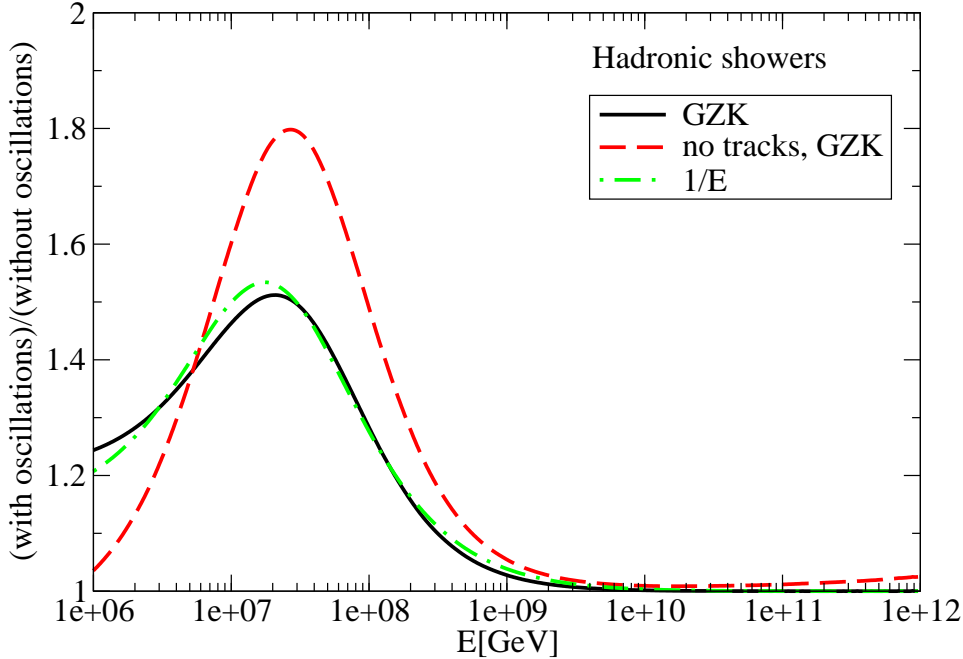}
\caption{Ratio of hadronic shower rates in the presence
and absence of 
$\nu_\mu\to\nu_\tau$ oscillations for GZK and $1/E$ fluxes 
for nadir angle $85^\circ$ for a km size detector.}
\label{fig:shratioh}
\end{figure}

Fig. \ref{fig:shratioh} shows the ratio of the hadronic shower distributions 
in presence or absence of oscillations for the same fluxes and nadir angle.
Without oscillations, the hadronic showers come from $\nu_e$ and $\nu_\mu$ 
interactions. In presence of oscillations $\nu_\tau$ and the $\nu_e$ and 
$\nu_\mu$ secondaries from tau decay contribute as 
described above. For hadronic showers the 
enhancement due to the presence of $\nu_\tau$ is $\sim$ 50\% at energies of a 
few $\times 10^7$ GeV. For hadrons there is also some small enhancement for 
shower energies between $10^8$ and $10^9$~GeV, which is not present for the 
electromagnetic showers. 

The tau contribution can be even more clearly seen if some of the other 
flavors can be separately identified. As previously discussed, IceCube
can identify showers from $\nu_\mu$ and $\nu_\tau$ charged current 
interactions by the $\mu$ or $\tau$ tracks that emerge and exit the detector. 
Once these can be removed, what remains are hadronic showers from
neutral current interactions of all flavors of neutrinos, 
$\nu_e$ CC interactions in the detector, as well as the hadronic showers 
from $\tau$ decays in the 
detector (both from taus produced within the
detector and from taus produced outside the detector that propagate in 
and then decay). The enhancement in the ratio is even higher in this case,
about $80\%$. 

\begin{figure}[t]
\epsfig{file=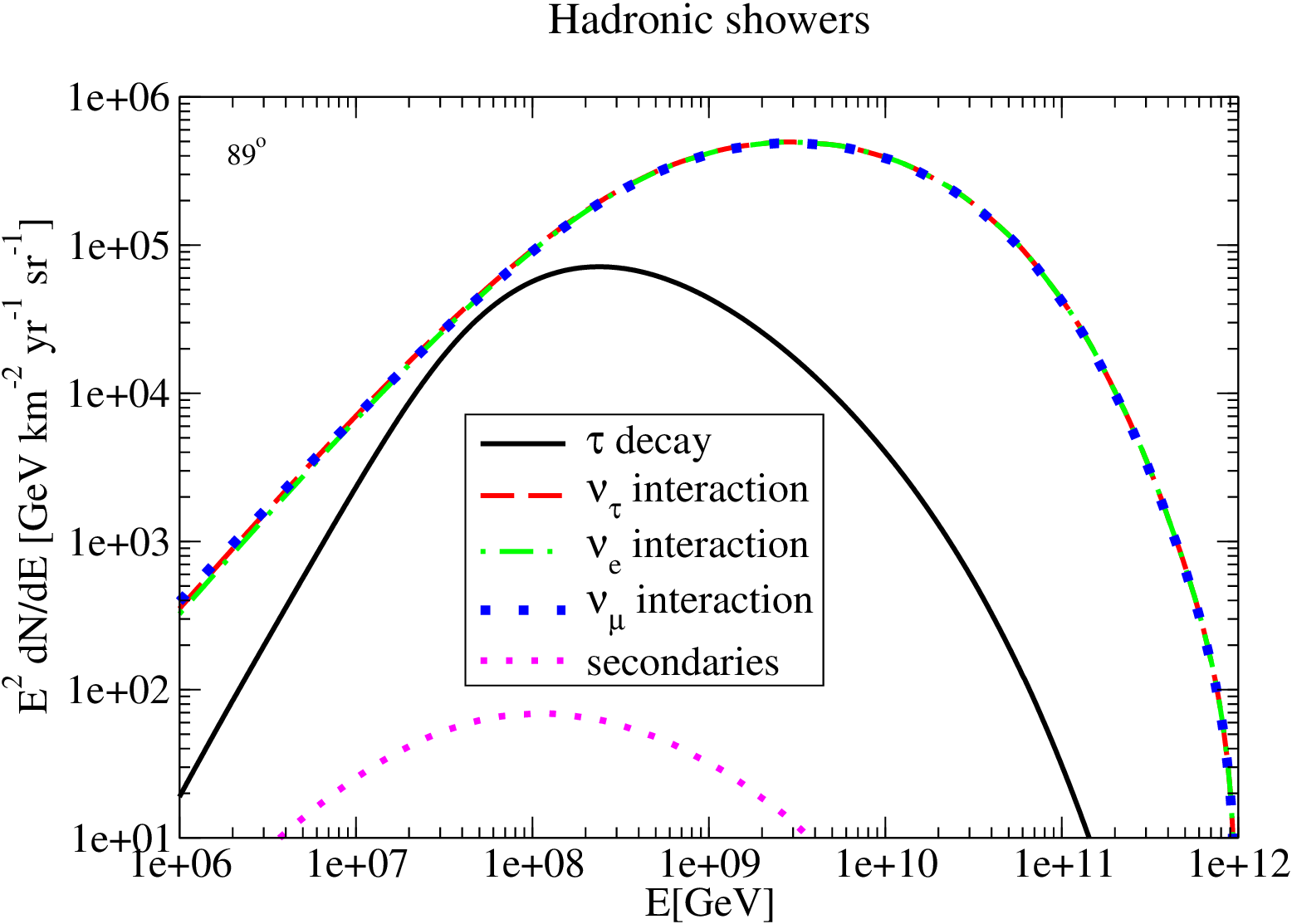,width=5in}
\caption{Hadronic shower distributions for GZK neutrinos, 
at a nadir angle of $89^\circ$ for a km size detector.}
\label{fig:hsh89}
\end{figure}

Fig.~\ref{fig:hsh89} also shows the hadronic shower distributions for
GZK neutrinos, but at a nadir angle of  $89^\circ$. As expected, there are 
more showers for this distance than for the longer one, but the relative 
contribution of the tau is smaller in this case. This is because both 
the attenuation and the regeneration effects are much smaller for smaller 
column densities.

The experimental angular resolution for high energy showers is about 
$10^\circ$ to $25^\circ$. Averaging over such angles the
effect of the pileup is somewhat reduced (to $\sim 10\%$ of the event rate), 
since total rates are dominated by the trajectories 
that go through a small amount of matter. However, rates can get significant 
enhancements at low energies where the regeneration from tau decays adds an 
important contribution even for longer trajectories. Consequently, an 
experiment with low energy threshold has a better chance of detecting the 
effect. IceCube has a low energy threshold and an energy resolution of 
$\sim 10\%$, so it is in a good position to look for pileup effects, given 
a high 
enough neutrino flux.

The same general features remain true for different initial fluxes.
However, the energy distributions are different for each and the
enhancement due to regeneration appears at somewhat different energies. The
regeneration effects are also smaller for the fluxes steeper at high energy. 
For example, from Figs. \ref{fig:shratioem} and  \ref{fig:shratioh} 
it can be seen 
that for the $1/E$ flux, predicted, for example, for $Z$ burst (ZB)
models \cite{sigl}, the pileup is slightly bigger, but at a 
lower energy than for the GZK neutrino flux. This is just as expected, 
because the GZK flux is steeper at higher energy, but does not
fall as rapidly at intermediate
energies.

\begin{figure}[t]
\epsfig{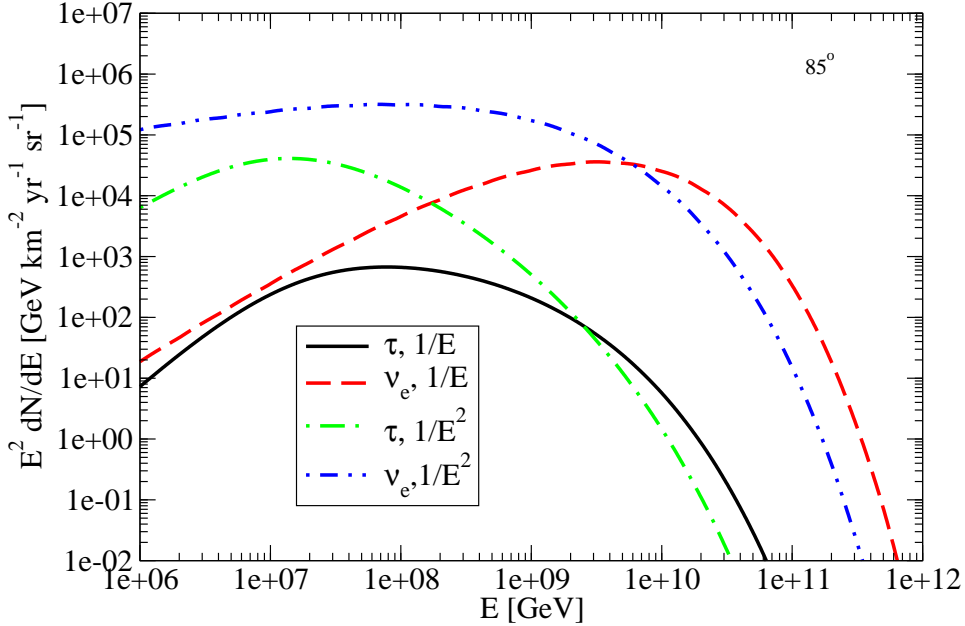}
\caption{Electromagnetic shower distributions
 for nadir angle $85^\circ$ for $1/E$ and $1/E^{2}$ characteristic fluxes 
for a km size detector from $\nu_e$ interactions and from $\tau$ decays.} 
\label{fig:emshdf}
\end{figure}

\begin{figure}[t]	
\epsfig{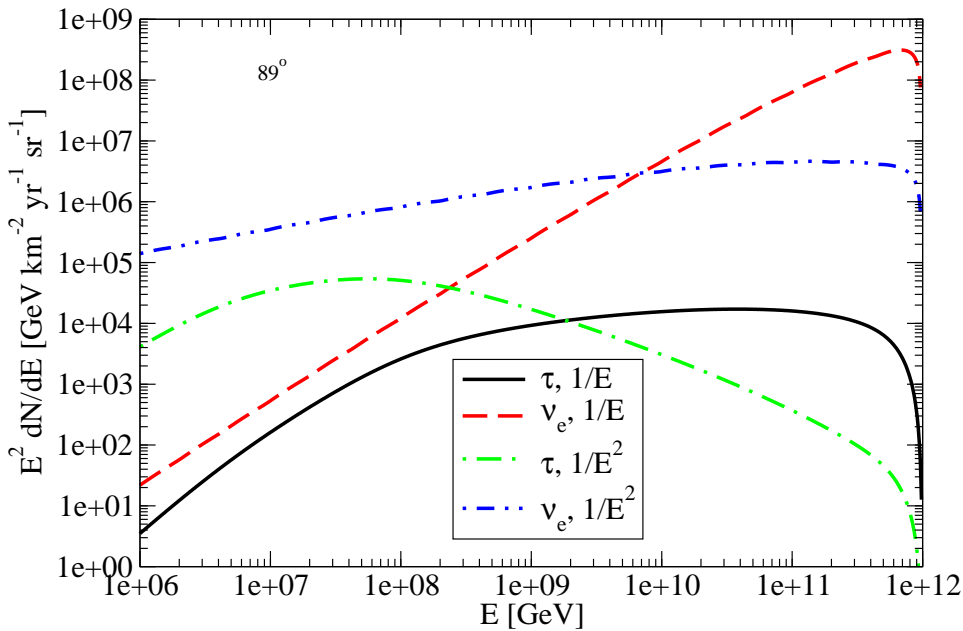}
\caption{Electromagnetic shower distributions
 for nadir angle $89^\circ$ for $1/E$ and $1/E^{2}$ characteristic fluxes 
for a km size detector.} 
\label{fig:emshdf89}
\end{figure}

Fig.~\ref{fig:emshdf} shows the  electromagnetic shower distributions for a 
$1/E$ flux, for example predicted for $Z$ burst and for 
a $1/E^2$ generic flux for a nadir angle of 
$85^\circ$. 
The $1/E^2$ distribution has a normalization of $10^{10}$ GeV$^{-1}$ 
km$^{-2}$ yr$^{-1}$ sr$^{-1}$, about an order of magnitude below the 
present AMANDA limit \cite{amanda}. 
For the ZB flux the normalization is 1 GeV$^{-1}$  km$^{-2}$ yr$^{-1}$ sr
$^{-1}$ 
up to $2.5\times10^{12}$ GeV. Above this energy the 
flux is cut off and drops as $1/E^3$. The 
steepness of the $1/E^2$ flux results in a small 
pileup and thus the 
relative contribution of taus is much smaller in this case. 
Fig.~\ref{fig:emshdf89} shows the same shower distributions for a nadir angle 
of $89^\circ$. Like for the GZK flux, the attenuation in this case is smaller
than for $85^\circ$, particularly at high energy. The effect of attenuation 
at high energy is striking especially for the $Z$ burst flux. At 
$85^\circ$ there is almost no flux left at energies above $10^{10}$ GeV, while
at $89^\circ$ the neutrino flux is almost unattenuated and still has the 
$1/E$ shape, being orders of magnitude higher than for the longer 
pathlength. However, the effects of regeneration become smaller for 
$89^\circ$ even for the less steep $Z$ burst flux. It can be clearly seen that,
depending on the energy threshold of the detectors, the contribution to 
event rates comes from different trajectories. At high energies the paths 
that go through the material with small column density will 
dominate event rates. However, at energies below $10^8$ GeV trajectories that 
go through 10-15 times more material can contribute equally due to the 
pileup and additional tau decays.

\begin{figure}[t]
\epsfig{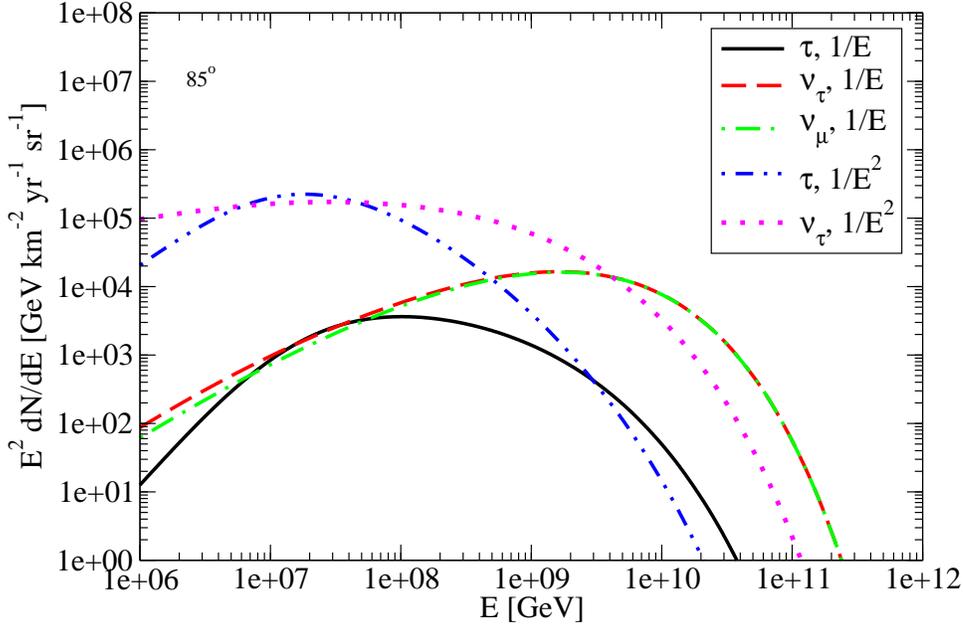}
\caption{Hadronic shower distributions 
for nadir angle $85^\circ$ for $1/E$ and $1/E^{2}$ characteristic fluxes 
for a km size detector from neutrino interactions and tau decays. The $\nu_e$ 
distributions are identical to the $\nu_\mu$ distributions. For the $1/E^2$
flux all neutrinos distributions are almost the same.} 
\label{fig:hshdf}
\end{figure}

Fig. \ref{fig:hshdf} 
shows the hadronic shower distributions for the fluxes used in
the previous figures, for a nadir angle of $85^\circ$. 
The energy distribution of the hadronic showers is different from that 
of electromagnetic ones due to the different decay distributions.
Taus decay mostly to hadrons and these hadrons carry most of the 
energy of the taus. Consequently, 
the tau contribution to hadronic showers is much
higher, bigger than the $\nu_\tau$ contribution at 
energies around a few $\times 10^7$ GeV, and extends to higher energies.

\begin{figure}[t]
\epsfig{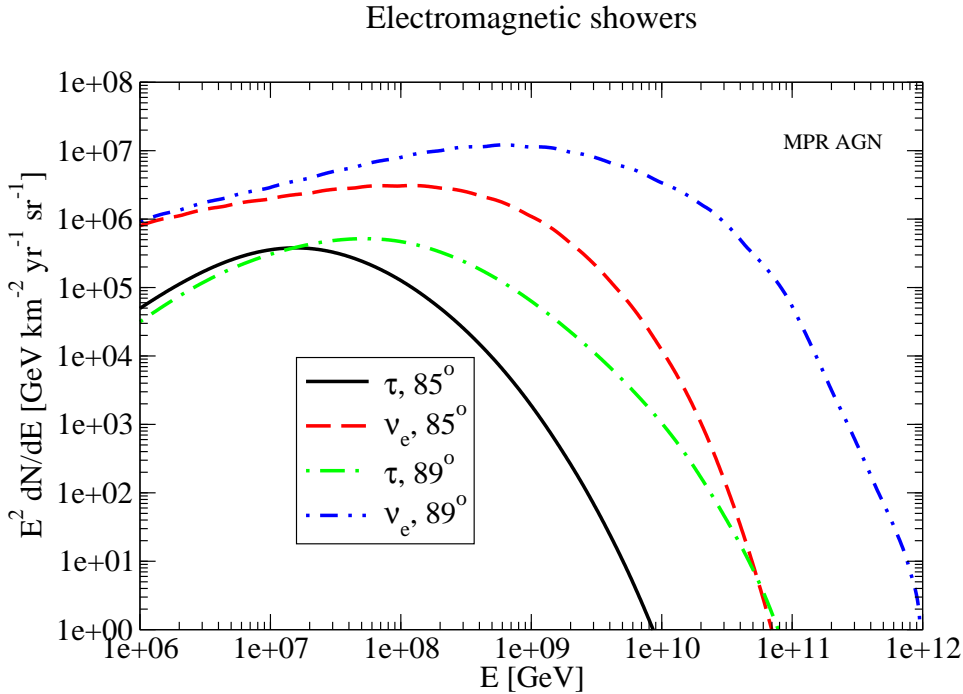}
\caption{Electromagnetic shower distributions
 for the MPR AGN model \cite{mpr} for a km size detector. } 
\label{fig:emshagn}
\end{figure}

Fig. \ref{fig:emshagn} shows the electromagnetic distributions for the 
AGN flux in \cite{mpr} for nadir angles of $85^\circ$ and $89^\circ$.
This flux is also steep at high energy and consequently the regeneration 
effects are small. The relative contribution of the taus is bigger 
for $85^\circ$, but the overall rates are higher for $89^\circ$.

The energy threshold of RICE is high and for the energies where the experiment 
has good sensitivity the $\nu_\tau$ enhancemnt is limited by the long
lifetime of the tau lepton and by its energy loss, such that it cannot be 
observed. We want to investigate if this remains true in the case of ANITA, 
which has a much larger effective volume and consequently has the potential
to detect many more tau's decaying over long trajectories.

\subsection{Showers in ANITA}

The ANITA experiment also uses the ice as a neutrino 
converter \cite{anita}. The long duration balloon missions will monitor 
the ice sheet from 40 km in altitude to a horizon approaching 700 km 
for refracted 
radio frequency signals with an effective telescope area of 1M km$^2$. The 
geometry of the experiment is rather complicated, as it has to take into 
account the Cherenkov angle of the radio emission with respect to the particle
trajectory, its refraction at the ice surface and the position of the balloon. 
The huge volume covered gives ANITA remarkable sensitivity for detecting 
very high energy neutrinos. While IceCube and RICE can only detect showers
produced in a ${\rm km}^3$ volume, ANITA can detect all 
the showers produced over long distances in ice. 

For example, for a nadir angle of $89^\circ$, the entire trajectory 
($ \sim 222\ {\rm km}$) of the neutrino is in ice, at less than 1 km 
depth, and all showers produced over this distance can be observed. 
For a nadir angle of $85^\circ$, observable showers (at less than 1 km depth) 
could be produced over a 
distance of $\sim$12 km. Given the very large detector area, a trajectory 
is not fully defined by the nadir angle of the incident neutrino at the 
entrance point, one also needs the position of the entrance/exit point with
respect to the icecap and the balloon. The trajectories that maximize the path 
through ice, rather than the ones that 
combine ice and rock, are the ones likely to dominate 
event rates because of the much smaller attenuation. Those going through more 
rock give larger {\it relative} 
enhancements due to regeneration. They could in principle contribute as much 
as the others in the region where the regeneration is effective.

As previously discussed, due to the showering over entire trajectories, we 
can no longer use Eq.~(\ref{emsht1}-\ref{nuhsh}), but rather we have to 
combine the propagation and showering from the beginning in order to 
correctly take into account neutral current interactions, energy loss of 
tau leptons and neutrino regeneration from tau decay.

\begin{figure}[t]
\epsfig{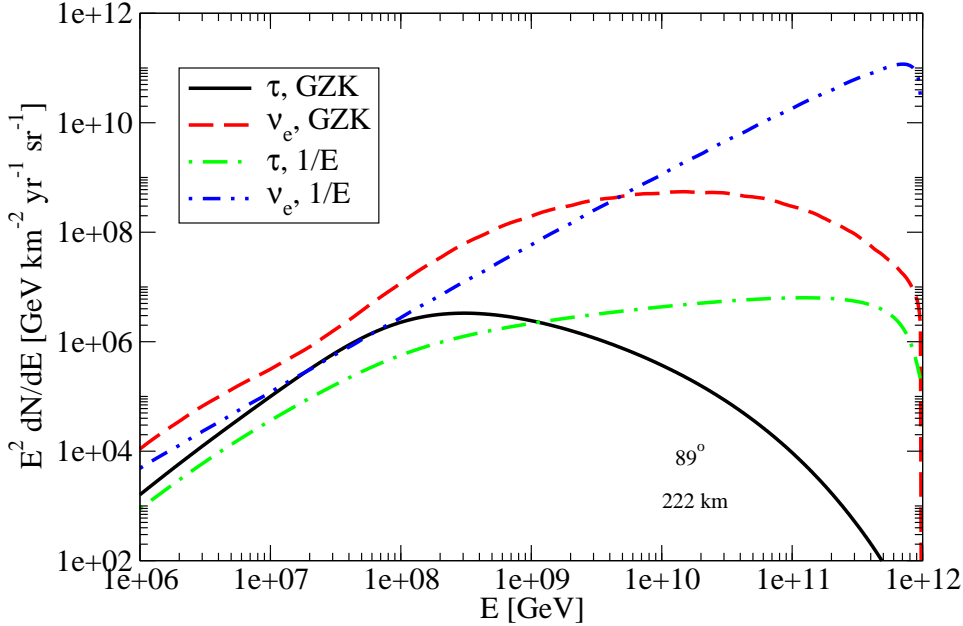}
\caption{Electromagnetic shower distributions for detection over 222 km of 
ice.}
\label{fig:emshanita}
\end{figure}

Fig. \ref{fig:emshanita} shows the electromagnetic shower distributions 
for the GZK and ZB initial fluxes over a trajectory of 222 km in ice.
Qualitatively, these are similar to the showers obtained in kilometer 
size detectors for the same trajectory. However, in this case the 
rates are larger by up to three orders of magnitude, depending on energy, 
due to the much larger surface of the detector. The energy distribution is 
somewhat different, being 
more spread out toward higher energies. Due to the longer 
detectable pathlength, higher energy taus, with decay length much longer 
than 1 km, 
can now decay producing observable showers.

\begin{figure}[t]
\epsfig{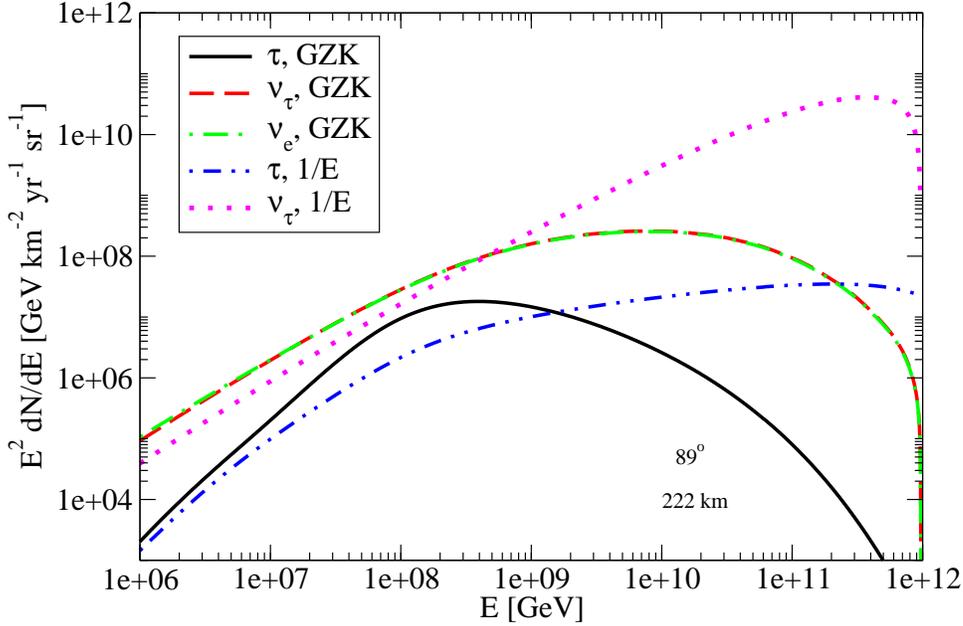}
\caption{Hadronic shower distributions for detection over 222 km of ice.
The $\nu_\mu$ distribution for GZK neutrinos is the same with that 
of $\nu_\tau$. For the $1/E$ flux all neutrino distributions are almost the 
same.}
\label{fig:hshanita}
\end{figure}

Fig. \ref{fig:hshanita} shows the hadronic shower distributions for the same 
initial fluxes and trajectory. The regeneration of the $\nu_\tau$ neutrinos 
in this case is very small and $\nu_\mu$ distributions are almost the same as
the $\nu_\tau$ ones. For the ZB flux, the $\nu_e$ distribution is the same as
for $\nu_\mu$, while for the GZK neutrinos it is different due to the 
difference in the initial fluxes.

\begin{figure}[ht]
\epsfig{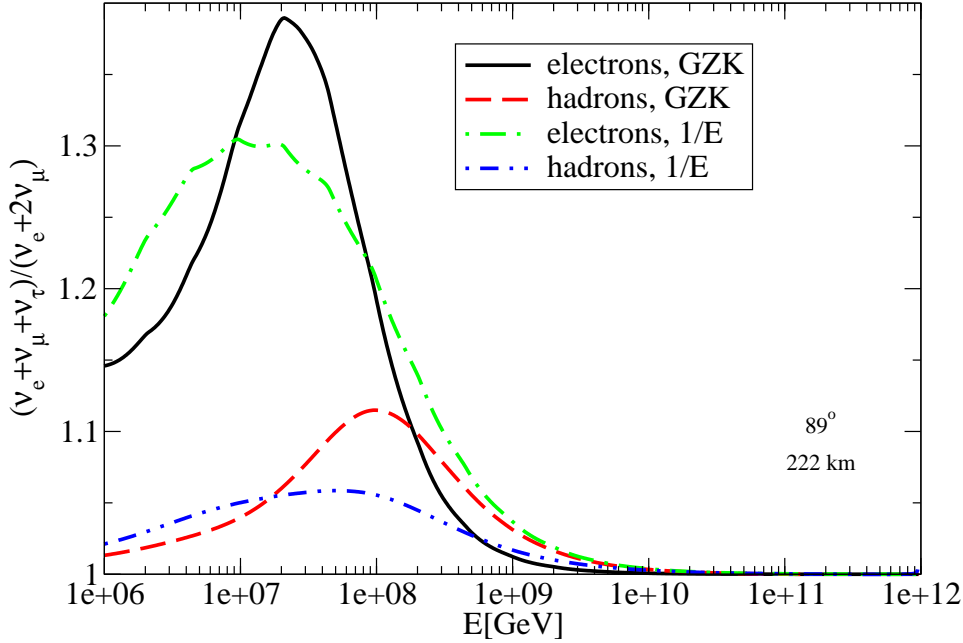}
\caption{Ratio of electromagnetic and hadronic shower rates in the presence
and absence of $\nu_\mu\to\nu_\tau$ oscillations for GZK  and 
$1/E$ fluxes of neutrinos
 for detection over 222 km of ice.}
\label{fig:shratioanita}
\end{figure}

Fig. \ref{fig:shratioanita} 
shows the ratio of electromagnetic and hadronic shower 
rates in the presence and absence of $\nu_\mu\to\nu_\tau$ oscillations 
corresponding to the distributions in the previous figures. The maximum 
enhancement due to the presence of $\nu_\tau$ is about 40\% at this angle, as 
expected since this trajectory has low column density. However, as previously
discussed, the enhancement occurs in a larger energy range and it peaks at 
higher energy than in the case of a small size detectors. In 1 km only taus 
with energies below a few $\times 10^7$ GeV have a significant probability 
to decay, while much higher energy taus can decay over the total distance of
more than 200 km.

\section{Conclusions}

We have studied in detail the propagation of all flavors of 
neutrinos with very high energy ($E \geq 10^6$ GeV) as 
they traverse the Earth.  Because of the high energies, we have limited
our consideration to nadir angles larger than $80^\circ$.
We are particularly interested in 
the contribution from tau neutrinos, 
produced in oscillations of extragalactic muon 
neutrinos as they travel large astrophysical distances.  
After propagation over
very long distances, neutrino oscillations change an initial (source) 
flavor ratio of $1:2:0$ to $1:1:1$ because of the
maximal $\nu_\mu\leftrightarrow\nu_\tau$ mixing. For GZK 
neutrinos, the flavor ratio at 
Earth 
deviates from 
1:1:1 because the incident fluxes are different. At lower energies and
smaller nadir angles, tau neutrino pileups from regeneration via
$\nu_\tau\rightarrow \tau\rightarrow \nu_\tau$ \cite{reg,reg1} enhanced
electromagnetic and hadronic signals in kilometer-sized detectors.
Our aim here was to see if there are similar effects at high energy.
  
In our propagation of neutrinos and charged leptons through the Earth, 
we have focused on 
kilometer-sized neutrino detectors, such as ICECUBE and RICE and 
on a detector with much larger effective area which uses Antarctic 
ice as a converter, ANITA.  Our study can easily be generalized 
to other experiments and to propagation in materials other than
ice.   

We have found that the $\nu_\tau$ 
flux above $10^8$ GeV resembles the $\nu_\mu$ flux. 
The lore that the Earth is transparent to tau neutrinos is not applicable
in the high energy regime. Tau neutrino pileups at small angles with respect to
the horizon are significantly damped due to tau electromagnetic energy loss above
$E_\tau\sim 10^8$ GeV if the column depth is at least as large as the neutrino
interaction length.

At lower energies, $E \leq 
10^8$ GeV, regeneration of $\nu_\tau$ becomes important for trajectories 
where the other flavors of neutrinos are strongly
attenuated. 
The regeneration effect depends strongly on the shape of the initial 
flux and it is larger for flatter fluxes.
The enhancement due to regeneration 
also depends on the amount of 
material traversed by neutrinos and leptons, i.e. on 
nadir 
angle. For GZK neutrinos, we have found that 
the enhancement peaks between $10^6$ and a few$\times 10^7$ GeV depending on
trajectory.
For $85^\circ$ 
the enhancement is 
about a factor of two, while for $80^\circ$ it 
is a factor of three.
  
We have translated the neutrino fluxes and tau lepton fluxes into rates for 
electromagnetic and hadronic showers at selected angles to see the effect
of attenuation, regeneration, and the different energy dependences of the
incident fluxes.  We have focused on comparing the $\nu_\tau$ contribution
to the $\nu_e$ and $\nu_\mu$ contributions to determine in what range, if any,
$\nu_\tau$'s enhance shower rates. 

The $\nu_\tau$ flux 
enhancements depend on the shape of the initial flux. The electromagnetic
showers are more sensitive to this shape than hadronic ones. The relative 
enhancement in hadronic showers is also smaller than for the electromagnetic 
showers. This is because for the electromagnetic signal the only contribution 
in the absence of taus is from electron neutrinos, while for hadrons the
tau contribution is compared to a much larger signal, from the interactions
of all flavors of neutrinos.
We have included contribution from secondary neutrinos, which 
we find to be relatively small for all fluxes.  

For kilometer-sized detectors, at for example a nadir angle of $85^\circ$, 
the maximal enhancemnet due to 
$\nu_\tau$ contribution to electromagnetic shower rates  
for the GZK flux is 
about $50\%$ at $3 \times 10^7$ GeV, while for the $1/E$ flux, it is even 
larger, 
about $70\%$, 
at slightly lower energy. In the case of hadronic showers for which the events
identified by muon tracks have been removed, the $\nu_\tau$ contribution 
peaks at about $2\times 10^7$ GeV and it gives an enhancement of about a 
factor of 1.8 for the GZK flux. These energy ranges are relevant for IceCube,
but not for RICE. For energies relevant to RICE,
tau neutrinos do not offer any appreciable
gain in electromagnetic shower signals compared to $\nu_e\rightarrow e$ CC
interactions, and they contribute at essentially the same level as $\nu_\mu$
to hadronic shower rates through NC interactions.

One of the reasons that tau neutrinos do not contribute large signals
to kilometer-sized detectors at very high energies is that high energy tau decay lengths
are very large, so the probability of a tau decaying in the detector is low.
For detectors like ANITA which can sample long trajectories through the
ice one would expect a larger tau neutrino contribution to the signal from 
tau decay.
Despite the long trajectory (222 km with a maximum depth of 1 km for a
neutrino incident at $89^\circ$ nadir angle) the tau contributions to
the electromagnetic shower rate is quite small for fluxes expected to contribute
in the ANITA signal. For hadronic showers, the suppression of $\tau$ decay to hadrons
relative to $\nu_e$ NC interaction contributions is about the same
as for electromagnetic showers compared to $\nu_e\rightarrow e$. The $\nu_\tau$ contribution to the
hadronic shower rate from interactions is the same as the $\nu_e$ contribution.
In summary, for ANITA, tau neutrinos do not give any additional signal beyond
what one would evaluate based on no regeneration from $\nu_\tau\rightarrow \tau
\rightarrow \nu_\tau$ due to tau electromagnetic energy loss at $E\aprge 10^8$ GeV. 
                
In addition to the experiments discussed here, there are many studies
concerning the possibility for detection of radio Cherenkov emission
from showers in materials other than ice.
It has been noted \cite{salsa} that rock salt formations have
similar properties to the Antarctic ice and can therefore be used as
large scale neutrino detectors. Salt has a higher density
($\rho_{salt}=2.2$ g/cm$^3$) than ice ($\rho_{ice}=0.9$ g/cm$^3$),
so it is possible to achieve an effective detection volume of several
hundred km$^3$ water equivalent in salt. This is somewhat larger 
than RICE, achieved with a much smaller actual detector size. 
The threshold for detecting the radio signal from showers in
salt is of the order of $\sim 10^{7}$ GeV, similar to RICE, but lower, such
that detection of extra signals
from $\nu_\tau$ enhancements would be more promising.

Also proposed is LOFAR \cite{lofar}, a digital telescope array
designed to detect radio Cherenkov emission in air showers. LOFAR has
sensitivity in an 
energy range of $\sim 10^{5} - 10^{11}$ GeV, so it can detect showers
at much lower energies than other radio Cherenkov experiments. 
LOFAR will likely be configured to detect horizontal showers from skimming
neutrinos as well. With its low energy threshold, LOFAR has an excellent 
opportunity to observe the shower enhancement at lower energies due to 
$\nu_{\tau}$ regeneration and tau pileup, which is not easily 
accessible in ANITA.

\section{Acknowledgments}

This research was supported in part by the National Science Foundation under
Grant No. PHY99-07949 and under DOE Contracts DE-FG02-91ER40664, 
DE-FG02-95ER40906 and DE-FG02-93ER40792. 
I.M. and I.S. thank KITP Santa Barbara for hospitality.
M.H.R. and I.S. thank the Aspen Center for Physics for
hospitality while this work was being completed.

\end{document}